\documentclass[preprint,showpacs,preprintnumbers,amsmath,amssymb,floats]{revtex4-1}
\usepackage{blindtext}
\usepackage{graphicx}
\usepackage[english]{babel}
\usepackage{amsmath}
\usepackage{amssymb}
\usepackage{color}

\newcommand{\be}{\begin{eqnarray}}
\newcommand{\ee}{\end{eqnarray}}
\usepackage[toc,page]{appendix}

\begin{document}
\title{Shot Noise near Quantum-Criticality}
\author{Srinivas Raghu}
\affiliation{Stanford Institute for Theoretical Physics, Stanford University, Stanford, CA. 94305}
\author{Chandra M. Varma}
\affiliation{University of California, Berkeley, CA. 94704\\
\thanks{Visiting Scholar} 
University of California, Riverside, CA. 92521} 
\thanks{Emeritus}
\date{\today}
\begin{abstract}
Shot-noise measures the correlations of fluctuations of current for a voltage applied much larger than the temperature and reveals aspects of correlations in fermions beyond those revealed in the conductivity. Recent measurements of shot-noise in the quantum-critical region of the heavy-fermion compound YbRh$_2$Si$_2$ (YRS) have presented a conceptual challenge to old theory and those devised following the experiments. 
Since the measured resistivity and the specific heat in YRS follow the predictions of marginal Fermi liquid (MFL) theory, we use it to calculate noise using the method developed by Nagaev. We get fair agreement with the magnitude and temperature dependence in the experiments using parameters from resistivity measurements. To achieve this, we find it necessary that the collisions between fermions by exchanging the MFL fluctuations conserve energy but lose momentum through Umklapp scattering and that the fermions and their fluctuations are locally in mutual equilibrium. 
%and that the self-energy rides the local chemical potential.   
At low temperatures, impurity scattering determines the noise and at high temperatures the MFL scattering. We show that  the noise for MFL scattering for high T alone is the same as the  Johnson-Nyquist noise, which in this case is temperature independent. Therefore  the Fano factor crosses over to $0$ at high  temperatures  independent of the voltage applied.
\end{abstract}
\maketitle

 {\it Introduction -} A variety of  quasi-2-dimensional metals have resistivity proportional to $T$ (apart from a constant offset due to impurity scattering) continuing down to  asymptotically low temperatures, together with a specific heat $\propto T \ln (\frac{\omega_c}{T})$ near  continuous $T=0$ phase transitions. These, as well as a variety of other properties \cite{LohneysenRMP2017, VarmaRMP2020}  have so far been understandable  as diverse manifestations of the marginal Fermi-liquid (MFL) phenomenology \cite{CMV-MFL}. The aim of this paper is to show that the recent measurements of shot noise and its temperature dependence \cite{Natelson2023} in
the anti-ferromagnetic quantum-critical region of the heavy-fermion compound YbRh$_2$Si$_2$ (YRS) \cite{Paschen2010}, with unexpected results, are qualitatively consistent with MFL phenomenology and put  constraints on microscopic origins of this phenomenology.  In particular, the scattering mechanisms responsible for momentum relaxation strange metallic transport, and those responsible for shot noise suppression are one and the same, and come from collective modes of the fermions.  These modes must therefore satisfy the same  local hydrodynamic constraints as the fermions themselves.

Shot-noise measures the zero-frequency limit of the correlations of current fluctuations due to a voltage induced energy $eV \gg k_BT_a$, the thermal energy at the ambient temperature $T_a$, in a quasi-linear geometry.  See Ref. \cite{BB2001} for a review.  For a two dimensional system of length $L$, width $W$, with $L \gg W$, the zero frequency shot-noise power is 
\be
\label{S1}
S_J(V,T_a) &=& 2 \int_0^{\infty} dt  \langle \delta I(t) \delta I(0) \rangle \nonumber\\
&=& 2 \frac{W^2}{L^2} \int_0^{\infty} dt \int_{-L/2}^{L/2} dx \int_{-L/2}^{L/2} dx' \langle \delta J(x,t) \delta J(x',0) \rangle  
\ee
where $\delta I$ is the instantaneous deviation of the  current from its steady state average  $\langle I \rangle$, with  the voltage difference $V$  between the contacts at $x=\pm L/2$, and $\delta J$ is the associated current density. 
%Since it is a measure of the correlations 
%in time of the granular charge traveling between two electrodes, it reveals the nature of the many-body correlations that %governs the current flow.
A fundamental issue on which the magnitude and temperature dependence of the calculated noise depends is whether the fluctuations that couple to fermions to give marginal Fermi-liquid are the fluctuations of the same fermions responsible for the electrical conduction,  or are they  of some other degrees of freedom. In the former case  the total energy of the fermions is conserved in the  inelastic collisions induced by the fluctuations. This is not so generally in the latter case. 

 The  classic calculations on shot-noise and the  Nyquist noise for metals not at criticality are easily summarized \cite{BB2001}. There is no shot-noise for ballistic propagation of charge. The noise due to Poisson thermal fluctuations at $T_a=0$  is
\be
\label{S-P}
S_{J}(V,0) = 2 e \langle I \rangle,
\ee
At zero applied voltage, the Johnson-Nyquist \cite{Nyquist} classical thermal noise  is 
$S_N(0, T_a) = 4 G T_a,$  
where $G$ is the conductance. 
%Note that for $T_a \gg  \tau_{i}^{-1}$, the Nyquist  noise for a MFL is a constant.
 
 It has become traditional to express noise by the ``Fano-factor" $F$ as the ratio of the noise minus the noise at  $V =0$ to  $2e\langle I \rangle$. 
\be
\label{F}
F(V,T) \equiv \frac{S_J(T_a,V) - S_J(T_a,0)}{2 e  \langle I \rangle}
\ee

{\it Sources of current relaxation -} Current correlations may occur due to  diverse physical processes besides the thermal noise. We shall not concern ourselves with noise when scattering by phonons is important  because that is irrelevant to the recent experiments. For current fluctuations due to elastic impurity scattering, calculations  in two opposite limits are available: 1) impurity scattering alone or 2) impurity scattering supplemented by  frequency independent particle-particle scattering (pps) in the strong scattering limit.  In the former, the shot-noise has been derived using transfer-matrix methods by Beenakker and Buttiker \cite{Beenakker1992} The same result was derived by Nagaev \cite{Nagaev1995, Rudin1995} using the Boltzmann equation with a Langevin noise source.  For  impurity scattering alone in the limit $T \to 0$,
\be
\label{1/3}
F_{elastic}(V, 0) = \frac{1}{3}.
\ee
 In the other limit, it is assumed that pps is strong enough that it leads to  local thermodynamic  equilibrium of the carriers on a length scale much smaller than the meso-scopic sample size in which the experiment is done, so that a local chemical potential $\mu(x)$ and a local temperature $T(x)$ can be defined. It is also assumed that pps conserves momentum so that it does not contribute to the resistivity. Moreover a frequency independent effective interaction of pps is considered. Under these assumptions, a relation using the Boltzmann equation is used to determine the relation between $T(x)$ and $\mu(x)$. The Fano-factor is calculated only at $T =0$ to be
$F_{strong}(V, 0) =\frac{\sqrt{3}}{4}$.
This result and its extension to finite $T$ or its Fermi-liquid version \cite{Si_Shotnoise2022} is insufficient in situations in which pps determines the temperature dependence of the low temperature resistivity.  This is the situation of interest to us both at criticality and the cross-over to the Fermi-liquid region, in which the resistivity changes to  $\rho_0 + A T^2$ and specific heat to $\propto T$ with strong numerical renormalization.  It is therefore required both that the pps be energy and temperature dependent and that it lose momentum through Umklapp scattering. 

 The scattering rate for any inelastic pps tends to $0$ as $T \to 0$ so that the equilibration length becomes larger than the sample length at low enough $T$.  The assumption of local equilibrium is then invalid. Therefore, in the $T \to 0$ limit, 
shot-noise reflects one of two possibilities.  First, if  elastic scattering from impurities dominates, the Fano factor $F \to 1/3$ is consistent with Eq. (\ref{1/3}).  Alternatively, if static screening of electron-electron interactions by the solid state environment is sufficiently strong, one can expect $F \to \sqrt{3}/4$ as $T \to 0$.   Both possibilities appear to have been  observed \cite{BB2001, Steinbach1996}. 
We will show below that the issue of lack of local equilibrium does not arise in the experimental conditions of Ref. \cite{Natelson2023} where the lowest measured temperature is $3 K$. 

As mentioned above,  at high enough temperatures, the inelastic pps of MFL dominates over the impurity scattering in determining the scattering rate  so that the conductivity is 
approximately proportional to $T^{-1}$. Then the Nyquist noise at $V =0$ is temperature independent.  We will also  show below that for a MFL alone  i.e. neglecting impurity scattering, the shot-noise is independent of $V$. It then follows that the $F$-factor defined by Eq. (\ref{F}) must go to $0$ at high enough voltage and temperature. We present below the calculation  of the cross-over from impurity dominated noise to MFL dominated noise. 
 Using estimates of the ratio of impurity scattering rate to the linear in $T$ scattering rate 
from the resistivity measurements, and of the voltage across the sample at different temperatures,  we find that our results are  in approximate agreement with the experiments. The qualifications to this success are
mentioned.

{\it Shot noise from the kinetic equation - }Our work rests on extensions of  the methods devised by Nagaev \cite{Nagaev1995, Rudin1995} to get Eqs. (\ref{1/3}) and $F_{strong}(V, 0)$ to include the physics of energy conserving  inelastic collisions induced by MFL fluctuations. These
 use the Boltzmann equation supplemented by the Langevin noise source as implemented by Kagan and Shulman \cite{Kogan1968}.  The steady state non-equilibrium Boltzmann equation for the distribution function $f(\bm r, \bm k, T)$ with a Langevin noise source ${\mathcal L}$ is
\begin{equation}
\left( \dot{\bm r}  \nabla_{\bm r} +  \dot{\bm k} \nabla_{\bm k} \right)f(\bm r, \bm k,T) = I_{coll} + {\mathcal L}.
\end{equation}
The collision integral is a sum of elastic scattering due to impurities and inelastic scattering from exchange by fermions of  fluctuations giving MFL.
\be
I_{coll} = (I_{el} + I_{inel}).
\ee
We consider a geometry as in  the experiment with a sample of length $L$ large compared to its transverse dimensions with a voltage $V$ across the two ends in the direction $\hat{x}$. The distribution function is assumed to depend only on $x$ and $\dot{\bf k} = e E \hat x$, $e E L = eV$. The  momentum dependence  of $f$ is written in terms of the dispersion of fermions and separated into two parts $f_{e} + f_{o}$, where $f_{e,o}$ is (even, odd) with respect to ${\bm k}$. Dropping the Langevin term for the moment, the Boltzmann equation then takes the form 
\begin{equation}
v_F \left(  \partial_x + eE_x \partial_{\epsilon} \right) (f_e(x, \epsilon,T) + f_o(x, \epsilon,T)) = I_{coll}.
\end{equation}
Since the energy is relative to the local chemical potential, the distribution function depends on the quantity $\epsilon +eEx$, and the two partial derivatives above can be expressed as the total derivative with respect to x \cite{Loss1998, BB2001}
\begin{equation}
v_F \frac{d}{dx} (f_e(x,\epsilon,T) +  f_0(x,\epsilon,T)) = I_{coll}.
\end{equation}
We can divide the collision integral into even and odd parts and write by symmetry that 
\begin{eqnarray}
v_F \frac{d}{dx} f_o(x, \epsilon, T) &=& I_{coll,e}(x, \epsilon, T) \nonumber \\
v_F \frac{d}{dx} f_e(x,y(x)) &=& I_{coll, o}(\epsilon, T)) = -\frac{1}{\tau(x,E,T)} f_o(x, \epsilon, T)
\end{eqnarray}
The last equality introduces the momentum relaxation time, and relates the odd and the even distribution functions:
\begin{equation}
f_o(x, \epsilon, T) = - v_F \tau(x,\epsilon, T_a) \frac{d}{dx} f_e(x,\epsilon, T_a))
\end{equation}
We therefore have
\be
\label{BE}
\frac{d}{dx}D(x, \epsilon, T) \frac{d}{dx} f_o(x, \epsilon, T) = I_{coll,e}(x, \epsilon, T).
\ee
where 
\be
D(x, \epsilon, T) = \frac{1}{2} v_F^2 \tau(x,\epsilon,T).
\ee
$\tau^{-1}(x,\epsilon,T)$ is the sum of the momentum relaxation rates due to impurities $\tau^{-1}_i + \tau^{-1}_{MFL}$, where in the latter the contribution due to Umklapp scattering  is included. Recently
the  transport properties (electrical and thermal conductivity, and thermo-power) \cite{Maebashi-Kubo} based on a microscopic theory of the Marginal Fermi-liquid \cite{Aji-V-qcf1, ZhuChenCMV2015, ZhuHouV} have been derived. It is shown that the transport scattering rate, necessary above, is the same as the single-particle scattering rate multiplied by a constant $< 1$, to account for vertex corrections due to Umklapp scattering. %It is only due to Umklapp processes that any pp scattering can give momentum loss.

It is not necessary for our purposes to  write down usual the inelastic collision integral $I_e$, (See for example \cite{Loss1998} - Eq. 227), for scattering of incoming fermions at energy and space ($\epsilon, x)$ and $(\epsilon',x)$ to outgoing fermions at ($\epsilon -\omega, x)$  and ($\epsilon' + \omega, x)$
through interactions with a Kernel $\chi"(\omega,T) = \chi_0 \tanh(\omega/2T)$, which is local in space, and has an upper cut-off $\omega_c$.
$I_e$ gives the single-particle relaxation rate. $I_0$, which gives the transport rate is $I_e$ multiplied by a factor $<1$ due to momentum loss through Umklapp scattering, as already mentioned.
 For later purposes we define a dimensionless coupling constant $\lambda \equiv g^2 \chi_0 N(0)$, where $\chi_0$ is the amplitude of the MFL fluctuations and $N(0)$ is the density of states of fermions near the chemical potential. $g$ is the renormalized fermion-collective mode vertex.

The collision integral as in the description above satisfies energy conservation in collisions through MFL fluctuations.  This must occur  if the fluctuations are collective modes of the same fermions that give the $T \ln T$ specific heat, do the conduction and give the conductance fluctuations. Then
\begin{equation}
\int_{-\infty}^{\infty} d\epsilon  (\epsilon - \mu) I_{coll,s}(\epsilon, E, x, T) = 0
\end{equation}
%It is easy to see that the diffusive left side of the Boltzmann Equation (\ref{BE}) integrated over $\epsilon$ is zero. So that the charge conservation is satisfied exactly. 
It is hard to solve the Boltzmann equation (\ref{BE}) exactly for the distribution function together with the local energy conservation condition. We use  the same strategy as Nagaev. We use an ansatz for the distribution function so that the local energy conservation condition on the collision integral is satisfied and use that on the left side of Eq. (\ref{BE}). Local thermodynamic equilibrium is equivalent to having a local chemical potential, 
$\mu(x) \equiv \mu - e E x$, 
and a local temperature $T(x)$. The local distribution function then is
\be
f_0(\epsilon, x, T) = \big(e^{\frac{\epsilon + E x - \mu}{T(x)}} +1\big)^{-1}.
\ee
One must  also consistently have an $x$ dependent  transport rate  because the fermions as well as the collective fluctuations are in local thermodynamic equilibrium in our theory. Specifically for a MFL with nearly momentum independent self-energy, where the compressibility is not renormalized \cite{Varma_HF_phen}, the self energy rides the chemical potential (in the present case, the local chemical potential) so that 
\be
\label{mfl-t}
\tau^{-1}_{MFL} = (\lambda |\epsilon + E x - \mu|) \coth \frac{|\epsilon + E x - \mu|}{2 T(x)}.
\ee

We determine the relation between $T(x)$ and $E x$ by weighting Eq. (\ref{BE}) by $(\epsilon -\mu)$  and integrating over $\epsilon$
\be
\label{Beq2}
\int_{-\infty}^{\infty} d\epsilon (\epsilon- \mu)  \frac{d}{dx} D_{tr} \frac{d}{dx}f_o = 0.
\ee
We have assumed that the energy $e V$ is much less than the bandwidth so that the limit on the integral can be taken to $\pm \infty$. Using that  the sample lies in the region $- L/2 \leqslant x \leqslant L/2$ and that in the middle at $x =0$, $dT/dx =0$, Eq. (\ref{Beq2}) gives
\be
\label{T1}
  \frac{dT^2}{dx} +  2 \frac{I_0(\lambda T(x), \tau_i)}{I_2(\lambda T(x) \tau_i)} E^2 x = 0.
\ee
Here
\be
I_n(\lambda T(x) \tau_i) \equiv  \int_{-\infty}^{\infty} dz z^n \frac{sech^2(z)}{1+  \tau_i \tau^{-1}_{mfl-tr}(z,x)}.
\ee
For  $\tau_i^{-1} \gg \lambda T(x)$, one gets Nagaev's result  $\frac{dT^2}{dx} = -2 \frac{I_0}{I_2} =-6/\pi^2 \approx - 0.608$ and for $ \lambda T(x) \gg \tau_i^{-1}$, $- 2\frac{I_0}{I_2} \to \frac{7\zeta(3)}{\pi^2} \approx -0.853$, independent of $T(x)$. We have checked that the result as a function of $\lambda \tau_i T(x)$ of importance for the experiments is close to the second limit. We will therefore consider only this limit which is realized over almost all of the range of $x$ for $eV \gg T_a$. With the boundary condition that $T(L/2) = T_a$, the ambient temperature, the solution is
\be
\label{T2}
T(x') = \sqrt{T^2_a + \frac{7\zeta(3)}{\pi^2} V^2 (1 - 4 x'^2)}, ~(x' = x/L).  
\ee
We may now re-insert the Langevin noise ${\mathcal L}$ in the Boltzmann equation to calculate $S$. Again, adapting  Nagaev's result for our case,  
\be
\label{S2}
S_I(V,T_a) =  4 \big(\frac{G(T)}{\tau^0_{tr}(T,\tau_i)} \big)  \int_{-1/2}^{1/2} dx'~ \int_{-\infty}^{\infty} d\epsilon ~\tau_{tr} (x',\epsilon) f_o(x',\epsilon)(1-f_o(x',\epsilon)).
\ee
We have introduced the factor in parenthesis in front of the integral to take care of all material dependent constants. $G(T)$ is the linear conductivity for the sample which is proportional to  $\tau^0_{tr}(T)$. 

 Eq. (\ref{S2}) is derived for normal fermions and is obviously valid in a Landau-Fermi-liquid. The question of its validity  at and near quantum-criticality where quasi-particles are not well-defined is interesting. We will consider marginal Fermi-liquids at quantum criticality,  in which the fermion self-energy has a smooth momentum dependence but  a weakly non-analytic frequency dependence; consequently, the Fermi-surface in momentum space remains well-defined. In that case, it has been shown that this is sufficient  for a calculation of steady state non-equilibrium processes such as the electrical and thermal conductivity and thermopower \cite{Maebashi-Kubo}. Actually the Boltzmann theory holds with only particle energies renormalized by the logarithmic inverse quasi-particle amplitude, which cancels due to a Ward identity for the conductivity.   This is also shown recently in a calculation of noise in \cite{Sachdevnoise2023}.

Under the conditions of  local equilibrium Eq. (\ref{S2}) used in Eq. (\ref{F}) gives,
\be
\label{F2}
F(V,T_a) &=& \frac{1}{V} (1+1/r_0)\Big( \int_{0}^{1/2} dx' ~ T(x') I_0(r(x')) -  T_a I_0(V=0) \Big), \\
I_0(r(x')) &\equiv & \int_{0}^{\infty} dz \frac{sech^2(z)}{1+ 2 r_0\frac{T(x)}{T_a}  z \coth(z)}.
\ee
where $r_0 \equiv 2 \lambda T_a \tau_i$. 
%For the pure limit $r \to \infty$, the result $F(V,T_a) = 0$ is directly obtained, as stated earlier.

{\it Comparison with experiment - } Let us now consider the region of applicability of Eqs. (\ref{S2}, \ref{F2}) to the actual experiment in YRS, i.e. consider the condition for $\ell_{mfl} =v_F \tau_{tr-mfl} \ll L$. $L$ is of $O(5 \times 10^3)~ \AA$ . It is not easy in a multi-band compound like YRS to extract $1/\tau_i$ and $\lambda T_a$ separately from the resistivity measurements \cite{Paschen2010}. We need the latter for judging the applicability.  If we take the heavy band velocity alone from the coefficient of linear in $T$ specific heat, $v_F$ for the compound YRS is of $O(50 Kelvin - \AA)$. Then $\ell_{mfl}(T)= L$ for $T \approx 10 mK$. If an estimate of the average velocity from all the bands is roughly estimated \cite{Paschen2010}, $\ell_{mfl}(T)= L$ for $T \approx 0.15 K$. The correct answer is in between these limits.  In either limits, the local equilibrium condition is valid in the experimental range which is between are $3 K$  and $10 K$.  In Eq. (\ref{F2}), we need to know only the ratio $r_0$,  which can be extracted from the ratio of resistivity at two different temperatures. Even for this purpose we run into the difficulty. While $(\lambda \tau_i)^{-1} = (10 \pm 2) K$  in the two-terminal resistance measurements for the device in which the shot-noise is measured \cite{Natelson2023}, the original YRS from which it is fabricated has $(1.5 \pm 0.5) K$, although it has a similar slope in temperature of the resistivity. Some contact resistance and boundary resistance in the shot-noise devise is certainly present. To compare with experiments, we will use $r_0$ in the range of both values estimated. 

In experiments voltage noise is measured with a current source. To get the applied voltage,
we consider the differential resistance up to the  current  of 80 $\mu$-Amps, which is the typical range of measurements. This gives $e V \approx 40, 50, 60, 75  K$ at $3, 5, 7, 10 K$ at which the experimental results are available \cite{Natelson2023}. 

\begin{figure}
 \begin{center}
 \includegraphics[width= 1.0\columnwidth]{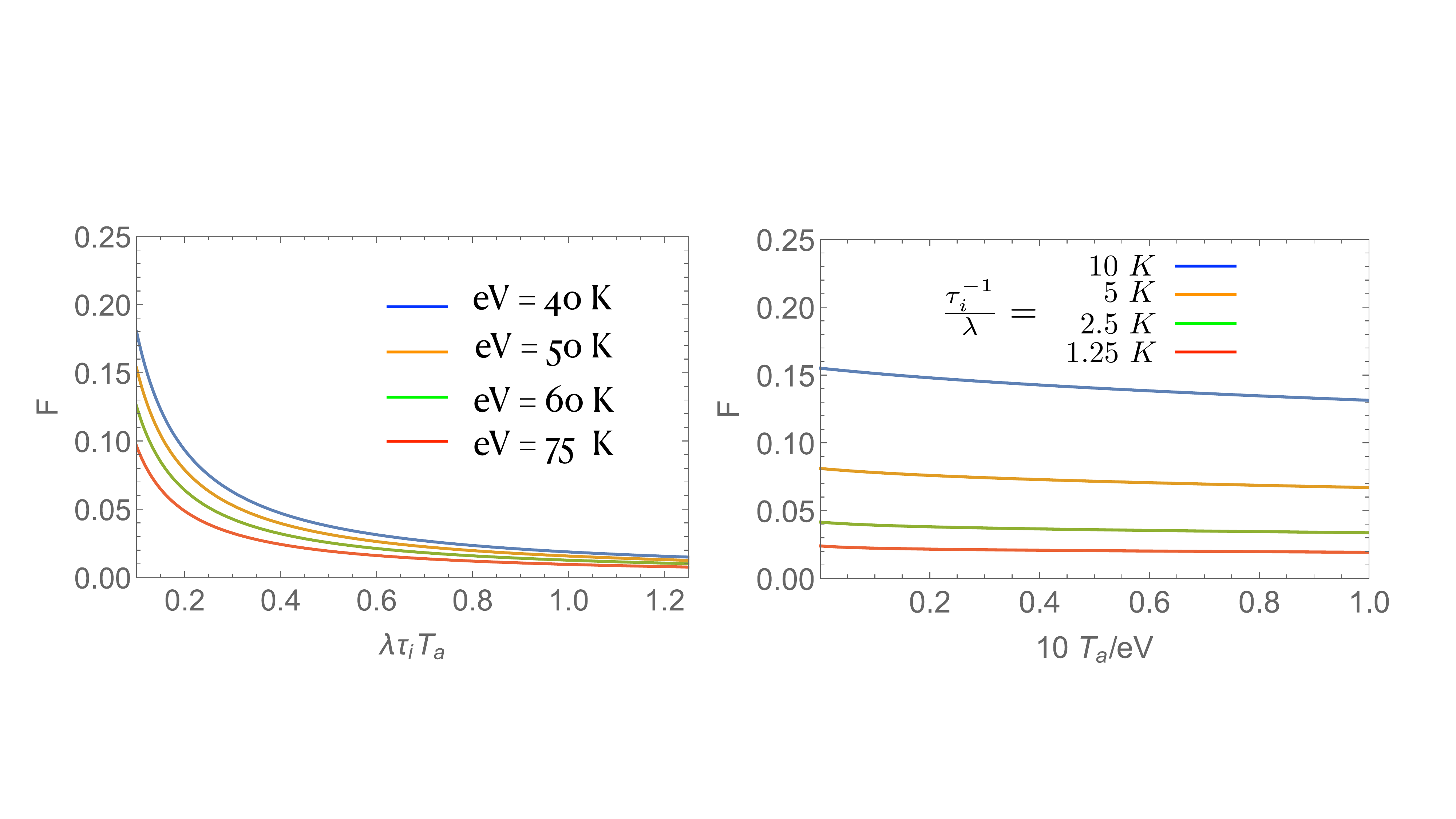}
 \end{center}
\caption{Left- Calculated $F$ as a function of $\lambda \tau_i T$ in the range of experiments and the four $eV$  estimated, one each for the four temperatures of measurements. Right- The same as a function of $T/eV$ at $eV = 50 K$ typical value for $\tau_i^{-1}/\lambda$. The conclusions are discussed in the text.} 
 \label{Fig:1}
\end{figure}

\begin{figure}
 \begin{center}
 \includegraphics[width= 1.0\columnwidth]{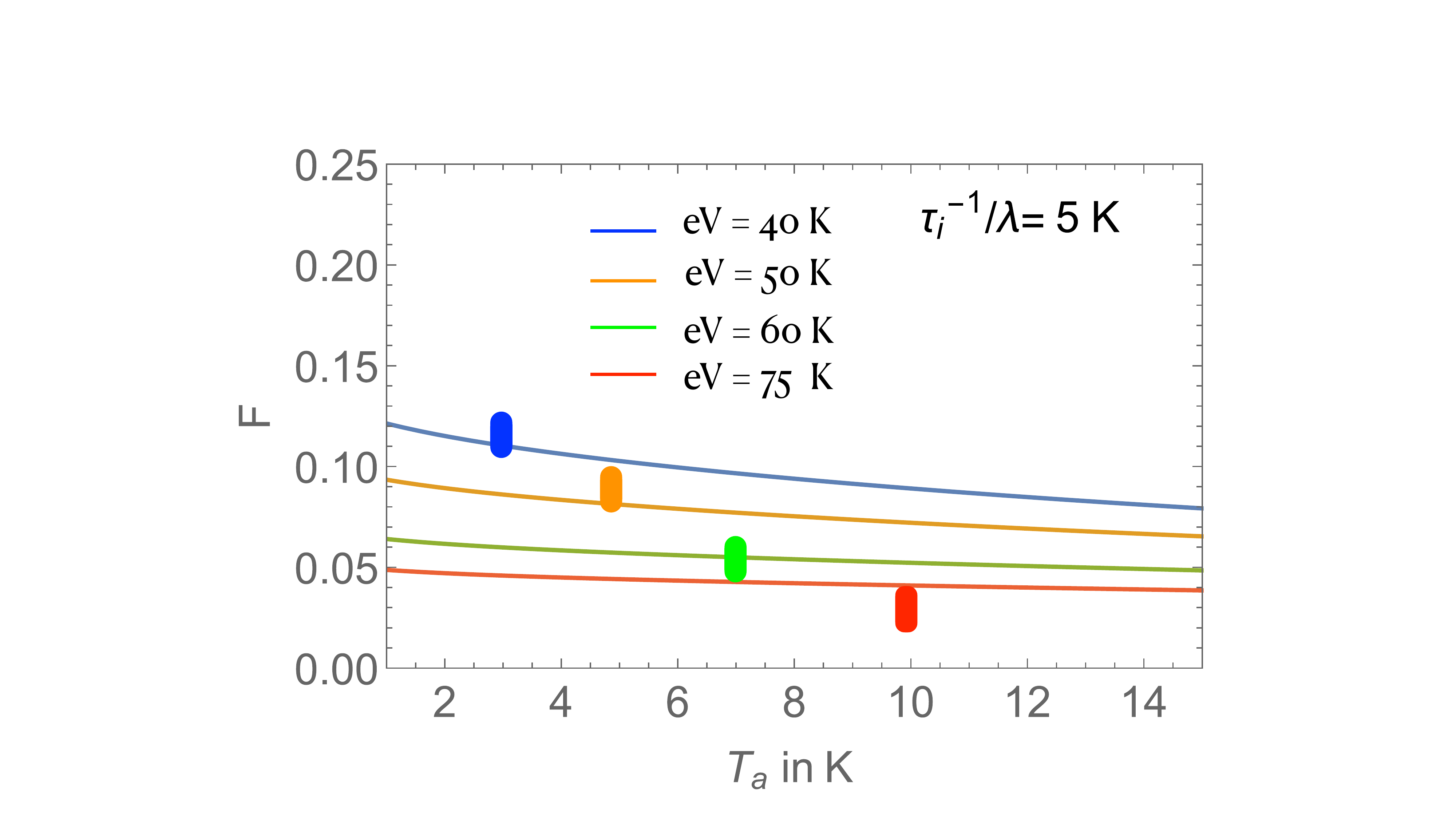}
 \end{center}
\caption{Comparison of the $F$ value at the four temperatures measured (3,5,7 and 10 K) with the calculated values plotted as a function of $T_a$ with the four values of $eV$ used in the experiment at those temperatures. The experimental points are to be compared with the calculation for the $eV$ estimated for that temperature and is color coded accordingly.}
 \label{Fig:2}
\end{figure}

We note that in the expression for $F$, $\tau_i$ always occurs in the combination $\lambda \tau_i T$, but $T_a$ occurs also weakly as $T_a/eV$ and as $T_a$ by itself. We show the calculated results in Fig. (\ref{Fig:1}) for $F$ as a function of $\lambda \tau_i T_a$ for the above values of $eV$. We also show in that figure the dependence on $T_a/eV$ for the specified values of $eV$ and a specific value of $\tau_i^{-1}/\lambda T$. These results give all features of $F$ for the relevant parameters, and accord with the qualitative conclusions given above. We note that the values of $F$ are in the range of experiments. For a fixed $T_a$, improved purity of the sample (increasing $\tau_i$) gives smaller $F$; larger values of $eV$ also suppress $F$ but more and more weakly as $V$ or $T$ increase. 

We can choose the value of $\lambda \tau_i^{-1}$ which gives the optimum comparison with the measured experimental results (Fig. \ref{Fig:2}) for $F$ as a function of $T_a$ for the values of $eV$ estimated. We also show their the four experimental points specified in the same colors as those for $eV$. The results are in fair agreement with experiments although one clearly discerns that the experimental data decreases somewhat faster than the calculations. The value for $\tau_i^{-1}/\lambda$ used is closer to that deduced in the samples used for the two-terminal experiments than for the very pure samples.  The lack of agreement with experiments at the highest $T$ may be because $eV$ which determines the local temperatures is above the Fermi-energy for the heavy mass particles. A test of the theory are low temperature measurements with varying $\tau_i^{-1}/\lambda$.

{\it Discussion - } This paper relies on collective fluctuations which have been derived for the quantum xy model coupling to fermions which have given the phenomenology of the MFL together with additional results \cite{Aji-V-qcf1, ZhuHouV}. The fluctuation spectra of YRS has not been measured but such experiments \cite{Schroder1, Schroder2} have been done on another heavy-fermion compound CeCu$_{6-x}$Au$_6$  with linear in T resistivity and specific heat proportional to $T \ln T$ near AFM quantum-criticality. The results are in agreement with the predictions of the microscopic theory as shown in detail in Ref. \cite{SchroderZhuV2015}.  We have shown within the context of a simple effective model that the low temperature linear in $T$ resistivity, and $T \ln T$ heat capacity are intimately tied with shot noise suppression.  
We predict similar shot noise suppression in all materials with quantum-critical regions with linear in T resistivity, like
the cuprates, other heavy-fermion materials  hosting AFM quantum-critical points, twisted bilayer graphene and WSe$_2$.
 
There have been recent works \cite{Sachdevnoise2023} \cite{Fosternoise2024}, which devise critical bosons of unspecified physical nature coupling to fermions together with varieties of hypothesized disorder to give linear in $T$ resistivity and $T \ln T$ specific heat. In calculating noise, it is assumed contrary to the requirement in our model that the effective temperature of such bosons is not the  same as that of the fermions. The calculations in these papers have not been compared with the  measured magnitude and temperature dependence of noise.

{\it Acknowledgements}: CMV thanks Douglas Natelson for very many informative discussions and explanations in relation to experiments and Lijun Zhu for lessons in calculating with Mathematica. SR is grateful to H. Goldman,  J.Yu and especially Y.-M. Wu for helpful discussions and collaboration on related topics.  We also thank  Qimiao Si for discussions on the theoretical status on the problem.   SR is supported  by the
US Department of Energy, Office of Basic Energy Sciences, Division of Materials Sciences and Engineering,
under Contract No. DE-AC02-76SF00515.

\bibliography{REF_dec2023}

\end{document}